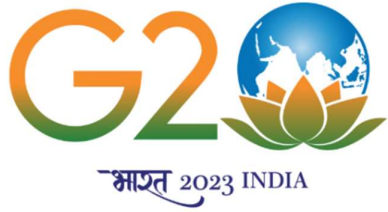 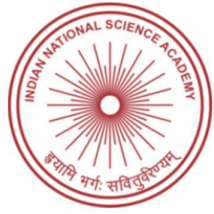 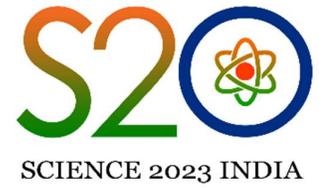

# Policy Brief

## GLOBAL DATA IN ASTRONOMY: CHALLENGES AND OPPORTUNITIES

Discussed and drafted during S20 Policy Webinar on Astroinformatics for Sustainable Development held on 6-7 July 2023

Contributors: Renée Hložek, Chenzhou Cui, Mark Allen, Patricia Whitelock, Jess McIver, Giuseppe Longo, Christopher Fluke, Ajit Kembhavi, Pranav Sharma, Ashish Mahabal

INDIAN NATIONAL SCIENCE ACADEMY – CENTRE FOR SCIENCE POLICY AND RESEARCH

**Introduction**
Astronomy is increasingly becoming a data-driven science. Advances in our understanding of the physical mechanisms at work in the Universe require building ever-more sensitive telescopes to gather observations of the cosmos to test and advance our theoretical models of how the universe works. To confront the observed data with our theoretical models we require data hosting, archiving and storage and high-performance computing resources to run the theoretical calculations and compare our simulated and observed universe. We also require the sophisticated development of highly skilled human resources.

New telescopes are coming online in the coming decade that will generate unprecedented amounts of data, and this will accelerate with telescopes like the Square Kilometer Array[1] and the Vera C. Rubin Observatory[2]. These telescopes are run through international collaborations and partnerships, driving a need for 'open science' and collaborative structure across national boundaries. While astronomical data are useful scientifically, the data do not come with the same ethical/privacy-related restrictions as medical/biological data. Moreover, the ability to use data for new scientific analysis extends and expands the impact and reach of scientific surveys – this is a strength that national funding agencies should capitalize on. However, the management and analysis of such large volumes of data present significant challenges that require policy-level preparations.

**Current Status of Global Data in Astronomy**
Ground-based telescopes, space observatories, and dedicated sky surveys have produced an enormous amount of data. Data products from these facilities include images, spectra, time series, data cubes, and catalogues of various types of celestial objects. While many astronomical data sets are open and accessible to the research community and also to the public, their discovery and use may require significant computational resources and technical expertise.

1. The landscape of telescope projects with large data volumes
   Our ability to survey the night sky has grown alongside the development of telescopes with larger apertures, the improvement in detector technology and the increase in speed of image-to-data conversion. The early 'mega-surveys' like the 1,500 square degree "Two-degree-Field Galaxy Redshift Survey (2dFGRS[3])" and the Sloan Digital Sky Survey (SDSS[4]), which has covered 15,000 square degrees and imaged a billion galaxies by its 18th data release, generated a wealth of data released to the public which enabled studies greatly beyond the planned scientific scope of the projects. The upcoming Vera C. Rubin Observatory, which will collect 20 TB of data per day and will optically image an estimated 38 billion objects over ten years, and the ambitious Square Kilometer Array project, which will collect an exabyte of raw radio data per day, compressed down to 10 PB/day, will generate more data than all previous telescopes combined. Processing, moving and curating these data will present one of the biggest challenges for the coming decades. The current Zwicky Transient Facility (ZTF[5]) is already demonstrating some of the complexities required in serving and processing "alerts" of new objects detected in its survey, which covers the night sky once every two days[6].

| Survey | Operation Date | Data volume (PB) |
|---|---:|---:|
| ZTF | 2017 | 1.5 |
| Gaia | 2018 | 2 |
| Pan-STARRS | 2019 | 1.4 |
| TESS | 2021 | 0.26 |
| Euclid | 2023 | 20 |
| Rubin/LSST | 2024 | 60 |
| Roman | 2027 | 20 |
| SKA | 2028 | 600 |

Table 1: Data volumes for a few current and future surveys

2. Consistent data formats and standards
   Astronomical data formats are often natively determined by the telescope system from which they are derived. The need for data storage, archiving and computation has been highlighted in the recent US planning process in the physics and astronomy communities[7,8]. As such, astronomy has faced the issue of bespoke data formatting and data type conversion for many years. In the early 2000s, a global movement towards consistent data formats and standards grew to form the International Virtual Observatory Alliance (IVOA), with member consortia including the US Virtual Astronomical Observatory (VAO) which grew out of the US National Virtual Observatory (NVO), the Astronomical Virtual Observatory (from the European Southern Observatory or ESO), Astronet, the UK AstroGrid and more[9]. While national funding for virtual observatory projects has fluctuated, the recommended IVOA standards still underpin many contemporary telescope data formats and standards, and the tools developed by the IVOA for data access and viewing are embedded into planned 'Science Platforms' of telescopes like the Vera C. Rubin Observatory. The need for support of open data has also been emphasized in other 'data-rich' physics fields[10].

---

[5] Bellm, E. et al The Zwicky Transient Facility: System Overview, Performance, and First Results PASP 131 018002 (2019)
[6] Masci, F. et al. The Zwicky Transient Facility: Data Processing, Products, and Archive, Publications of the Astronomical Society of the Pacific, 131:018003 (30pp), 2019
[7] Alvarez et al. Data Preservation for Cosmology Submitted to the Proceedings of the US Community Study on the Future of Particle Physics Available online at https://snowmass21.org/submissions/cf (2021)
[8] Bailey, S. et al. Data and Analysis Preservation, Recasting, and Reinterpretation Submitted to the Proceedings of the US Community Study on the Future of Particle Physics Available online at https://snowmass21.org/submissions/cf (2021)
[9] R. Hanisch et al. The Virtual Astronomical Observatory: Re-engineering access to astronomical data Volume 11, Part B, June 2015, Pages 190-209
[10] Nachman, B. When, Where, and How to Open Data: A Personal Perspective Proceedings of the US Community Study on the Future of Particle Physics Available online at https://snowmass21.org/submissions/cf (2021)

3. Public data successes: a case example in the Sloan Digital Sky Survey
   The SDSS was a revolutionary telescope in that it was able to photometrically image the sky rapidly, leading to increased co-added image depth relative to previous surveys, and the ability to detect transients and variable stars on a nightly basis. Szalay described how the computational archiving and storage of the survey led to a large return on the initial investment of the survey and phrased the problem in terms of 'funding dollars': For the SDSS, the cost was roughly USD 500k/year to maintain the data servers. Given that a scientific paper can be costed at roughly USD 100k/year (if papers from funded projects take roughly a year to generate): only 5 papers per year from SDSS data would justify the archiving and storage needs. To date, roughly 11,000 papers mention the SDSS and its successors – illustrating the impact of storage and archiving of data.

**Challenges and Opportunities**

Some of the major challenges include data management, data curation, data quality control, data access, and data sharing. Intercontinental Internet bandwidth is an important bottleneck to global astronomical data sharing. While recognizing the value of Science Platforms and interoperable services in addressing these challenges, the computational cost of conducting large-scale data joint analysis based on Science Platforms brings new challenges. Therefore, we need for policies that would promote standardization, interoperability, and integration of data from different sources along with the development of pipelines and provenance tools to ensure reproducibility and transparency of astronomical research.

The FAIR principles (that data should be Findable, Accessible, Interoperable, and Reusable)[11] were particularly valuable when applied in the astronomical context. Policies that enable and facilitate data sharing are the key to progress in this area, and that Astronomy can be a great way to test new ideas/models for sharing data given the limited need for security/privacy protocols on astronomical data must be stressed. The biggest challenge is ensuring that funding and infrastructure grow to match the policy/strategic vision of participating countries. Data storage and processing are themselves infrastructure costs that can amount to ~5-10% of the operational budget of a large telescope, but these costs are often left out of the operational budget and planning process. There is also a mixed ability of nations to enforce open data principles in scientific research: Data Management Plans are sometimes called for in proposals but are rarely enforced. While some tools and services exist to facilitate open access (e.g., certified data centres and methods for minting DOIs), building tools and infrastructure to guide FAIR data and code will lead to further improvements in this area: this is where G20 leadership can have an outsized impact on this field.

The opportunities presented by global data in astronomy include new discoveries and insights into the nature of the universe, better understanding of the formation and evolution of celestial objects, and contributions to global development goals for quality education as well as for Industry, innovation and infrastructure, enhancing the public trust in science via transparent access and engagement. Initiatives under the International Astronomical Union's Office of Astronomy for Development (OAD), which encompass projects aimed at catalysing economic growth, enhancing diplomatic relations, and facilitating the transfer of knowledge and expertise from astronomy to society, provide a compelling illustration of the potential for constructive convergence between these domains.

---

[11] Wilkinson, M. et al. The FAIR Guiding Principles for scientific data management and stewardship. Sci Data 3, 160018 (2016)

**Recommendations**

The S20 Academies should assist their governments in:
1. Developing guidelines for open sharing of astroinformatics data from nationally funded scientific projects, while ensuring data quality, security, and privacy.
2. Sharing best practice adopting standards for data formats, metadata, and interfaces to promote interoperability and integration of data from different sources.
3. Adoption of highly successful IVOA standards for astronomical data.
4. Sharing best practices for open access of code and make resolutions to adhere to FAIR principles.
5. Investing in data management and curation tools, including pipelines and provenance tools, to ensure reproducibility and transparency of astronomical research.
6. Improving the global cyber-infrastructure for scientific research, including Astronomy.
7. Developing a strategy to promote data sharing and collaboration among astronomers, data scientists, and other stakeholders.
8. Sharing best practice for payment policy and solutions for data converged use in Science Platforms.
9. Developing capacity-building training programs in data management, curation, and analysis at various levels, from professional astronomers and STEM practitioners to students and learners

**S20 Co-Chair**: Ashutosh Sharma, Indian National Science Academy
**INSA S20 Coordination Chair:** Narinder Mehra, Indian National Science Academy

**Contributors**
Renée Hložek, University of Toronto, Canada
Chenzhou Cui, National Astronomical Observatories, China
Mark Allen, Centre de Données astronomiques de Strasbourg, France
Patricia Whitelock, South African Astronomical Observatory, South Africa
Jess McIver, University of British Columbia, Canada
Giuseppe Longo, Università degli Studi di Napoli Federico II, Italy
Christopher Fluke, Swinburne University of Technology, Australia
Ajit Kembhavi, Inter-University Centre for Astronomy and Astrophysics, India
Pranav Sharma, Indian National Science Academy, India
Ashish Mahabal, California Institute of Technology, USA